\documentstyle[12pt,aasms4]{article}

\received{29 May 1998}
\accepted{3 August 1998}
\lefthead{Lutz et al.}
\righthead{Mid-Infrared Spectroscopic Survey of ULIRGs}

\begin{document}

\title{The nature and evolution of Ultraluminous Infrared Galaxies:
  A mid-infrared spectroscopic survey
   \footnote{Based on observations with ISO, an
   ESA project with instruments funded by ESA Member States (especially the PI
   countries: France, Germany, the Netherlands and the United Kingdom) with the 
   participation of ISAS and NASA}
 }
\author{D. Lutz\footnote{Max-Planck-Institut f\"ur extraterrestrische Physik, 
   Postfach 1603, 85740 Garching, Germany},
  H.W.W. Spoon$^2$,
  D. Rigopoulou$^2$, 
  A.F.M. Moorwood\footnote{European Southern Observatory,
  Karl-Schwarzschild-Stra\ss\/e 1, 85748 Garching, Germany},
  R.Genzel$^2$
  }

\begin{abstract}
We report the first results of a low resolution mid-infrared
spectroscopic survey
of an unbiased, far-infrared selected sample of 60 ultraluminous infrared 
galaxies (ULIRGs, $L_{IR} > 10^{12} L_{\sun}$) 
using ISOPHOT-S on board ISO, the Infrared
Space Observatory. 

We use the ratio of the 7.7$\mu$m `PAH' emission feature to the local
continuum as a discriminator between starburst and AGN activity. About
80\% of all the ULIRGs are found to be predominantly powered by star
formation but the fraction of AGN powered objects increases with
luminosity. Whereas  only about 15\% of ULIRGs at luminosities below $2\times
10^{12} L_{\sun}$ are AGN powered this fraction increases to
about half at higher luminosity.

Observed ratios of the PAH features in ULIRGs differ slightly from those
in lower luminosity starbursts. This can be plausibly explained by the
higher extinction and/or different physical conditions in the
interstellar medium of ULIRGs.

The PAH feature-to-continuum ratio is anticorrelated with the ratio of
feature-free 5.9$\mu$m continuum to the IRAS 60$\mu$m continuum,
confirming suggestions that strong mid-IR continuum is a prime AGN
signature.  The location of starburst-dominated ULIRGs in such a
diagram is consistent with previous ISO-SWS spectroscopy which implies
significant extinction even in the mid-infrared. 

We have searched for indications that ULIRGs which are advanced
mergers might be more AGN-like, as postulated by the classical
evolutionary scenario. No such trend has been found amongst
those objects for which near infrared images are available to assess
their likely merger status.
 
\end{abstract}


\section{Introduction}

The nature of ultraluminous infrared galaxies (ULIRGs, see Sanders \&
Mirabel 1996\markcite{sanders96} for a recent review) has been the
subject of lively debate since their discovery by IRAS more than a
decade ago. Although evidence for both starburst and AGN activity in
ULIRGs has accumulated during this period, the question as to which
generally dominates the luminosity has remained largely unsolved,
mainly due to observational difficulties associated with their
large dust obscuration.

With the advent of the Infrared Space Observatory (ISO) of the
European Space Agency, sensitive mid-infrared spectroscopy became
available as a new tool capable of penetrating the obscuring
dust. Fine structure line and PAH feature observations with SWS and
ISOPHOT-S of a sample of 15 bright ULIRGs suggest that most are
starburst-powered (\cite{genzel98}). However, this sample is too small
to search for luminosity or evolutionary effects. Using only PHT-S it
has subsequently proved possible to extend to fainter sources and
increase to 60 the number of ULIRGs observed over the wavelength range
containing the 6.2, 7.7, 8.6, and 11.3$\mu$m features commonly
attributed to polycyclic aromatic hydrocarbons (PAH). Groundbased
observations of these and a companion at 3.3$\mu$m first demonstrated
that these features are strong in starburst galaxies but weak or
absent in classical AGNs (\cite{moorwood86}, \cite{roche91}). ISO
spectroscopy has further strengthened this link by demonstrating the
anti-correlation between feature strength relative to the continuum
and the ionization state of the gas (\cite{genzel98}). In the analysis of
the larger sample presented here, therefore, it is considered
reasonable to use the line to continuum ratio of the most prominent,
7.7$\mu$m, feature as our primary discriminator between starburst and
AGN activity.

The sample is drawn from the 1.2\,Jy survey
(\cite{fisher95,strauss92}).  We selected ULIRGs with
$L_{FIR}>10^{11.7} L_{\sun}$ (approximately equivalent to
$L_{IR}>10^{12} L_{\sun}$)\footnote{See e.g.  Sanders \& Mirabel
(1996) for the definition of the commonly used 40-120$\mu$m
($L_{FIR}$) and 8-1000$\mu$m ($L_{IR}$) luminosities}, $S_{60}>1.3Jy$;
good visibility for ISO; redshift below 0.3 and a clear optical
identification. No infrared color criteria were applied to avoid
biasing the sample in AGN content.  Our sample includes the ULIRGs of
Genzel et al. (1998)\markcite{genzel98} except NGC 6240 and IRAS
23060+0505 which do not meet the selection criteria.
We have adopted IRAS FSC fluxes
for all our sources to compute total (8-1000$\mu$m) IR luminosities
$L_{IR}$ (H$_0$=75, q$_0$=0.5), replacing flux upper limits by
estimates based on the 60\,$\mu$m flux and average far-infrared colors
of the ULIRGs of Sanders et al. (1988) where necessary.
\markcite{sanders88} Sources meeting our basic luminosity and flux
criteria may be missing because of poor ISO visibility, lack of good
optical identifications or because they were not scheduled during the
ISO mission. These effects do not, however,
affect the unbiased nature of the sample for studying the power
sources and evolution of ULIRGs.

\section{Observations and Data Analysis}

ISOPHOT-S 3-11.6$\mu$m spectra of the ULIRGs were obtained in chopped
mode, using pure on-source integration times between 512 and 2048
seconds.  ISOPHOT-S comprises two simultaneously operating grating
spectrometers.  Here, we only make use of the long wavelength section
covering the wavelength range 5.84 to 11.62$\mu$m because the S/N
ratio for the faint ULIRGs is too low in the short wavelength
section. The data processing was performed using standard procedures
of the PHT Interactive Analysis (PIA\footnote{PIA is a joint
development by the ESA Astrophysics Division and the ISOPHOT
Consortium led by the Max-Planck-Institut f\"ur Astronomie (MPIA)
Heidelberg.  Contributing ISOPHOT Consortium Institutes are DIAS, RAL,
AIP, MPIK, and MPIA}) software versions 6 and 7. The detector drift
modelling available in the most recent versions of PIA is problematic
for extremely faint sources and was therefore not used. Instead an
upward flux correction of 40\% to the not drift corrected data was made to
approximately take into account the drift effects (U. Klaas, private
communication).

\placefigure{fig:pah_meth}

The average of all 60 ULIRG spectra, individually scaled to
S$_{60}$=1Jy to give all sources equal weight
(Fig.~\ref{fig:pah_meth}), clearly shows the PAH features at 6.2, 7.7,
and 8.6$\mu$m but relatively weak continuum. Comparison with the
starburst and AGN templates provides a first and direct indication
that ULIRGs are, on average, starburst-like.  Fig.~\ref{fig:pah_meth}
also illustrates our method for extracting PAH and continuum data from
the individual spectra. Two pivots were used to determine the
continuum by linear interpolation - a feature-free continuum point at
5.9$\mu$m rest wavelength, shortward of the 6.2-8.6$\mu$m PAH complex
and the underlying emission plateau, and a continuum point at
10.9$\mu$m rest wavelength in the long wavelength flank of the silicate
absorption, but short of the 11.3$\mu$m PAH emission feature. In the relatively
frequent cases where  the 10.9$\mu$m point is redshifted out of the
ISOPHOT-S range we adopted $S_{10.9}=(2.5\pm0.5)\times S_{5.9}$ which
was found adequate for low redshift sources. In a few cases, the
linear continuum was estimated by eye because application of the
standard method gave unphysical results, e.g. interpolated 7$\mu$m
continuum above the observed spectrum. 

Since ULIRGs are heavily obscured,
silicate absorption may strongly influence the spectra beyond 8$\mu$m and
affect the continuum placement.
Because of flanking emission features and the ISOPHOT-S 
long wavelength cutoff, the silicate optical depth is poorly constrained
from our ISOPHOT data alone. To illustrate the effect of 
extinction, we have added as dashed lines in the top panels of 
Fig.~\ref{fig:pah_meth} the starburst and AGN templates additionally 
obscured by A$_V$=50 assuming the ISO-derived
extinction law of Lutz et al. (1996,1997). Obscured starbursts can still be
separated from obscured AGNs by presence of the 7.7 and 6.2$\mu$m PAH 
features. Both normal and obscured spectra demonstrate that the ratio between
narrow-band fluxes at 7.7$\mu$m (feature plus continuum) to 5.9$\mu$m 
(continuum) is a valuable discriminator between starbursts (high ratios) 
and AGNs, with little sensitivity to extinction since both points are outside 
the silicate feature. Analysing this ratio for our sample gives AGN/starburst 
discrimination closely similar to that obtained from the line-to-continuum 
method. Below, we only present results of the line-to-continuum method which 
places firmer limits on PAHs on a rising AGN continuum.  
Due to the proximity of 7.7$\mu$m to the reliable 5.9$\mu$m continuum point, 
errors in continuum placement are overall small enough not to affect the
discrimination between starbursts and AGNs. Heavily obscured AGN continua can 
mimick a 7.7$\mu$m PAH peak in this method but stay at L/C$\lesssim$1. In 
addition, obscured AGN would not show the 6.2$\mu$m feature that is clearly
seen in the average ULIRG spectrum.

Because the shape of PAH features is known to vary with physical conditions
(e.g. \cite{roelfsema98}), we did not attempt to fit a (fixed) PAH
shape to the data to measure the PAH features in our low S/N data. The
peak strength of the 7.7$\mu$m feature was measured by simply taking
the average of all data within a window covering the rest wavelengths
7.57--7.94$\mu$m. Errors or limits on all these quantities are
based on the noise measured shortward of 5.9$\mu$m rest wavelength and
the increase in noise towards longer wavelengths corresponding to the
decrease of the ISOPHOT-S spectral response.

\section{Starbursts vs. AGNs}

\placefigure{fig:pah_lum}

Figure~\ref{fig:pah_lum} shows the measured line-to-continuum ratios
(L/C) of the 7.7$\mu$m PAH feature plotted as a function of integrated
infrared luminosity. For reference, our average starburst and AGN
template spectra yield L/C=3 and 0.2 respectively. In the following
discussion we have therefore classified ULIRGs with L/C below 1 ( plus
some with upper limits slightly above 1) as AGNs and the rest as
starbursts. Only the average has been plotted for 14 ULIRGs which were
not individually detected reliably in either the feature or
continuum. The fact that these sources have weak 5.9$\mu$m continua
compared to those at  60$\mu$m (see below) indicates that these
are also powered by starbursts. This is supported by the average of
their spectra which, although noisy, shows a clear 7.7$\mu$m
feature. Note also that a selection effect in favour of detecting AGNs
is caused by the fact that, for the same 60$\mu$m continuum, an
average AGN will exhibit a 7.7$\mu$m continuum flux density which is
higher than the peak PAH flux density of a starburst
(Fig~\ref{fig:pah_meth}). The noisy spectra will hence preferentially
correspond to starburst-like ULIRGs. 

The distribution seen in Fig.~\ref{fig:pah_lum} suggests that the AGN
fraction rises towards high ULIRG luminosities. Dividing the sample
into two bins separated at log ($L_{IR}$)=12.3 and using the above
criteria we find most probable and (firm upper limit) AGN fractions of
8/51 (21/51) for the low luminosity bin, 5/9 (6/9) for the high and
13/60 (27/60) in total. The most probable values assume that the 14
individually undetected sources are starburst powered, for the reasons
given above, while the firm upper limits assume that they are all
AGNs. 

The above results provide both (i) a better statistical basis for the
conclusion, based on SWS and ISOPHOT-S spectroscopy of a smaller
sample, that most ULIRGs are predominantly starburst powered
(\cite{genzel98}) and (ii) support for the increase in AGN powered
fraction with luminosity inferred from optical spectroscopy
(\cite{veilleux97}). With regard to the latter, however, the AGN
fraction at log ($L_{IR}$)$<$12.3  estimated with the extinction-insensitive
PAH method is
still lower and, together with more detailed case studies such as that
for UGC\,5101 (\cite{genzel98}), cautions that some ULIRGs exhibiting
optical Seyfert spectra may still be predominantly starburst-powered.

\subsection{Ratios of PAH features}
Our average ULIRG spectrum (Fig.~\ref{fig:pah_meth}) exhibits a ratio
of the 6.2/7.7$\mu$m features which is slightly lower than in typical
starbursts. The measured ratio of the integrated feature fluxes is
0.25 compared to 0.30 for the average of the reference starburst
galaxies. This behaviour is more pronounced in some individual
spectra, e.g. for Arp 220 where the ratio is 0.18. One possibility is
that the weakness of the 6.2$\mu$m feature reflects the unusual
conditions of the ULIRG interstellar medium. ISO observations of
galactic sources may provide clues on such trends. For
compact \ion{H}{2} regions, Roelfsema et al. (1996, 1998)
\markcite{roelfsema96, roelfsema98} find a 6.2/7.7 ratio which on
average is lower than for `typical' \ion{H}{2} regions. Very low
6.2/7.7 ratios, lower than for the ULIRGs, are observed in some of
these compact \ion{H}{2} regions (Fig. 5 of \cite{roelfsema98}), as
well as high 8.6/7.7 ratios in others which might be linked to an
intense radiation field. A decrease of the 6.2/7.7$\mu$m ratio is also
observed going from the disk to the central starburst of  M\,82
(\cite{tran98}).

A second effect which could be responsible for unusual PAH ratios is
the influence of strong extinction which is already suggested by the
similarity of the average ULIRG spectrum and the obscured starburst spectrum
of Figure~\ref{fig:pah_meth}. Extinction suppresses the 6.2, 8.6,
and 11.3 features in comparison to the one at 7.7$\mu$m. This effect
will be even stronger than for a standard extinction curve if the
material obscuring the ULIRGs contains icy grains. Absorption features
due to water ice and other ices are observed with ISO in the 6$\mu$m
region and between the two silicate features in galactic sources such as
young stellar objects (e.g. \cite{whittet96}) and, to a lesser degree,
towards the center of our Galaxy (\cite{lutz96}) which samples a more
normal line of sight containing  some molecular material.  Of the starburst
templates with ISOPHOT-S PAH spectra and for which Genzel et
al. (1998) estimated extinction from independent ISO-SWS spectroscopy,
extinction approaches ULIRG levels only for NGC\,4945 and the
molecular ring encircling the center of our Galaxy. Interestingly,
these are the only spectra in that group which also show low
6.2/7.7$\mu$m flux ratios. Further, as in our average ULIRG spectrum,
their 8.6$\mu$m features appear as a shoulder to the 7.7$\mu$m rather
than as a separate feature, due to suppression by silicate absorption.
Attempts to estimate the strength of the PAH emission from the
11.3$\mu$m feature (e.g. \cite{dudley97} for Arp220) may underestimate
the importance of PAHs in highly obscured sources. This, in
consequence, will lead to an overestimate of the depth of the silicate
feature and problems in interpretation of the mid-infrared spectrum in
general.

\subsection{Dust, Destruction and Dilution?}

\placefigure{fig:pah_59}

More insight into the properties of ULIRGs can be gained from a
diagnostic diagram which combines the PAH line-to-continuum ratio with
the ratio of feature-free 5.9$\mu$m continuum (from our spectra) to
IRAS 60$\mu$m continuum (Fig.~\ref{fig:pah_59}). Strong 5.9$\mu$m
continuum appears in the case of AGNs due to the presence of hot dust
in the NLR or torus. We also show in this diagram the median location
of the small sample of known starbursts observed in addition to the
ULIRGs. Seyfert galaxies and QSOs scatter over a wider range of the
diagram, possibly due to varying contributions of star formation and
are less easily condensed into a median location. The AGN location
plotted is meant to represent `pure' AGNs and QSOs but note that most
of the PAH L/C ratios are only upper limits.  We have also plotted the
change vectors expected due to various effects. Dust extinction suppresses the
5.9$\mu$m continuum with respect to the far-infrared but does not affect
the PAH L/C ratio. Destruction of the PAH
carriers shifts a source vertically while dilution by an
AGN-powered hot dust continuum moves a source along a diagonal,
provided the PAH flux and the far-infrared continuum are unaffected.

A first result is the clear anticorrelation between PAH L/C ratio and
5.9 to 60$\mu$m flux ratio. A similar anticorrelation is found when
the 25 to 60$\mu$m ratio is used instead. Our ISO
PAH-spectrophotometry thus strongly indicates that `warm' ULIRGs (in
the sense of 5.9/60 or 25/60 flux ratios) are AGN dominated, while
`cold' ULIRGs are starburst dominated, in agreement with earlier
continuum-based conclusions (\cite{sanders88a}).

A second result is that those ULIRGs which are starburst-like in their
PAH L/C tend to have lower $S_{5.9}/S_{60}$ than the template starbursts.
We interpret this as the result of very high extinction in the
dust-rich ULIRGs, in agreement with our conclusion from the PAH feature
ratios. The corresponding 6$\mu$m extinctions of
up to 2.5 magnitudes (for a foreground screen) are consistent with
ULIRG 25$\mu$m screen extinctions of $\lesssim$1~mag derived from
\ion{H}{2} region emission lines (\cite{genzel98}) and the mid-IR
extinction curve of Lutz et al. (1996,1997)\markcite{lutz96}\markcite{lutz97}. 
Also, the
corresponding silicate feature optical depths of up to $\sim$5 agree with 
significant silicate optical depth indicated in Arp220 (\cite{smith89},
\cite{charmandaris97})
which interestingly is the source of lowest S$_{5.9}$/S$_{60}$ in
Figure~\ref{fig:pah_59}. Even
higher obscuration, that would hide the center completely in the
mid-infrared, has been inferred by Fischer et
al. (1997)\markcite{fischer97} from a fit to the shape of the ISO-LWS
far-infrared continuum of Arp 220. Adopting a fully mixed, single
temperature emitting slab model they obtain optical depth one (through
the complete slab) at 150$\mu$m, while optical depth one occurs near
40-50$\mu$m in mixed models equivalent to the screen extinctions we
derive above. Excellent fits to the
Arp 220 SED can also be achieved by assuming $\tau=1$ in the
30-50$\mu$m region and a mix of dust temperatures.  CO fluxes and size
limits, and FIR/submm continuum size limits (\cite{scoville97}) have
been used to infer that ULIRGS may be optically thick in the
far-infrared
(e.g. \cite{solomon97}). Differences between high CO-estimated optical
depth and actual optical depth to the starburst are also observed, however,
to the central regions of starburst galaxies like M82 and are attributed to a
separation between the bulk of the gas and star forming regions. A
similar separation may occur in Arp 220 where much of the gas appears to be
concentrated into a thin disk (\cite{scoville97}).

Third, no clear correlation or anticorrelation between the
ratio of PAH to FIR continuum flux and the 5.9/60$\mu$m flux ratio
can be seen in close inspection of Fig.~\ref{fig:pah_59}. 
Our ULIRG data are therefore consistent with a model in which
the PAH to far infrared emission is constant, mid-infrared extinction
is a few magnitudes and the AGNs simply contribute a diluting hot dust
continuum. This interpretation may, however, not be unique since some
template AGN which are not highly obscured (e.g. NGC 4151) are
similarly offset from the dotted line in Fig.~\ref{fig:pah_59}
representing pure dilution by hot dust.  This indicates {\em less} PAH
compared to the 60$\mu$m continuum than starbursts resulting e.g. from
destruction of their carriers.  It is hence equally possible that the 
extinction to at least some
AGN-like ULIRGs is substantially lower, and that the PAH emission is
not only diluted by continuum but indeed also suppressed to some
degree. Due to the diluting continuum, it is difficult to lower the
limits on PAH flux sufficiently to confirm this destruction, even for
bright sources with high S/N spectra. For illustration,
Fig.~\ref{fig:pah_59} also contains a mixing line between starburst
and AGN templates, labeled by varying starburst contribution to the
60$\mu$m flux. Note that this also indicates that choosing PAH L/C=1
to separate starbursts from AGN is conservative in the sense of not
misclassifying AGN as starbursts.

\section{Evolution of ULIRGs}

\placefigure{fig:pah_sep}

The classical evolutionary scenario of Sanders et
al. (1988)\markcite{sanders88} postulates that interaction and merging
of the ULIRG parent galaxies triggers starburst activity which later
subsides while the AGN increasingly dominates the luminosity and
expels the obscuring dust. An implication of this scenario would be
that advanced mergers should, on average, be more AGN-like than
earlier stages of interacting galaxies which are still well separated.
To test this, we have plotted in Fig.~\ref{fig:pah_sep} the PAH
line-to-continuum ratio as a function of the nuclear separations
estimated from various near-infrared imaging studies (\cite{armus94},
\cite{carico90}, \cite{duc97}, \cite{graham90}, \cite{majewski93},
\cite{murphy96}). Although near infrared imaging of our sample is
incomplete, it is already evident in this figure that starbursts are
found to the smallest nuclear separations and that there is no obvious
trend towards AGN dominance with decreasing separation. This suggests
that the dominance of AGN or starburst may depend on local
and shorter term conditions determining the fuelling of the AGN in addition
to the global state of the merger.

\acknowledgments
We are grateful to Eckhard Sturm, Michele Thornley and Dan Tran for discussions.
SWS and the ISO Spectrometer 
Data Center at MPE are supported by DLR (DARA) under grants 50 QI 8610 8 and 
50 QI 9402 3.

\clearpage
\figcaption[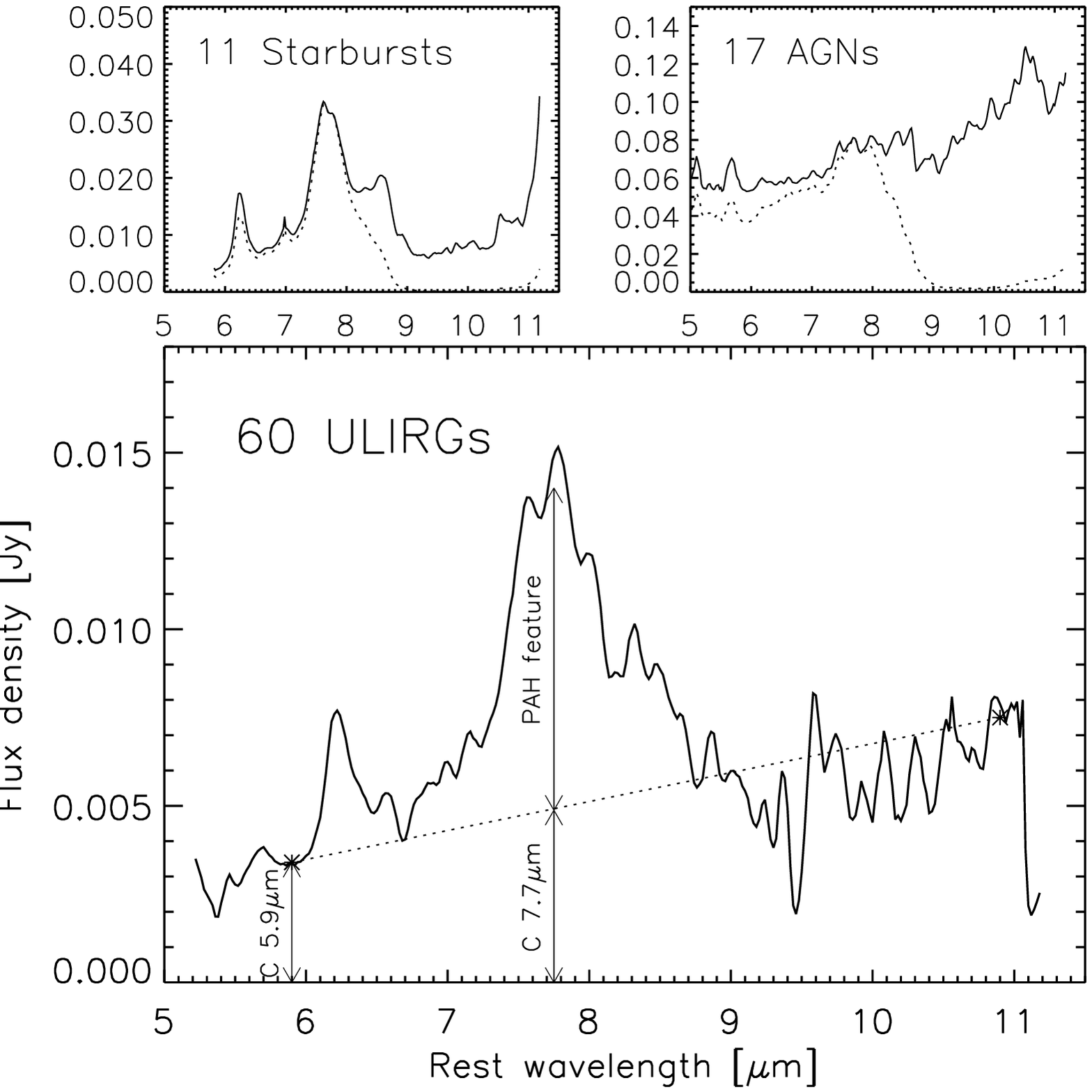]{Average ISOPHOT-S spectrum of all ULIRGs observed,
individually scaled to $S_{60}$=1Jy. The method used to derive the
line-to-continuum for the 7.7$\mu$m PAH feature, and the feature-free
5.9$\mu$m continuum is indicated (see also text). Average spectra of
starburst galaxies and AGNs are added for comparison. The dashed lines
represent these spectra after applying an additional A$_V$=50 foreground 
extinction and re-scaling for display purposes.
\label{fig:pah_meth}}

\figcaption[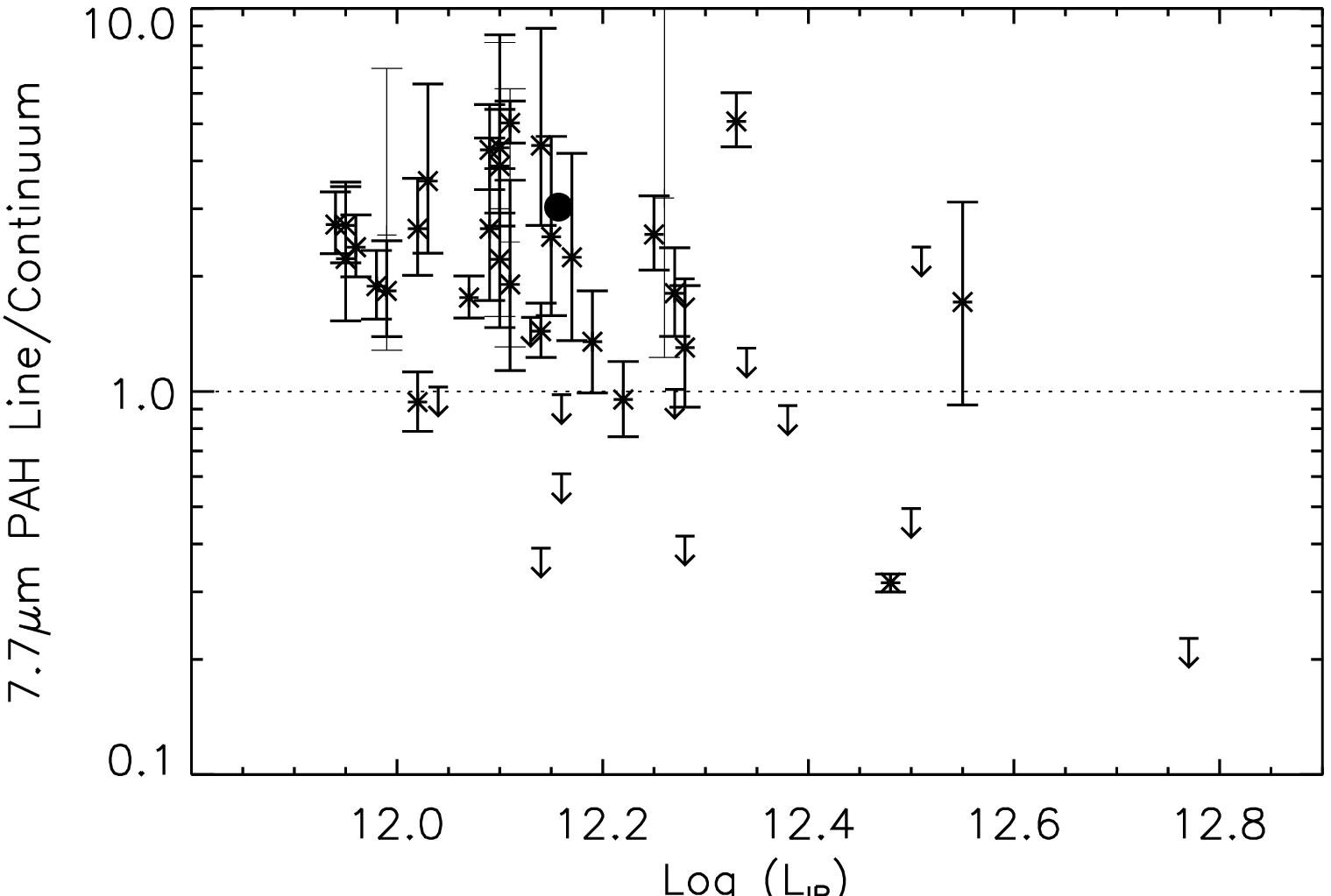]{Line-to-continuum ratio for the 7.7$\mu$m PAH 
feature as a function of infrared luminosity. Thick symbols refer to PAH 
detections or 3$\sigma$ limits
with reliable ($>2\sigma$) continuum detections. Thin symbols represent
galaxies with significant PAH but continuum detected at less than 2$\sigma$.
The filled circle gives the location of the {\em average} spectrum of 14 ULIRGs
that were too faint to be analysed individually. The dotted line indicates 
the adopted separation between starbursts and AGNs at L/C=1. 
\label{fig:pah_lum}}

\figcaption[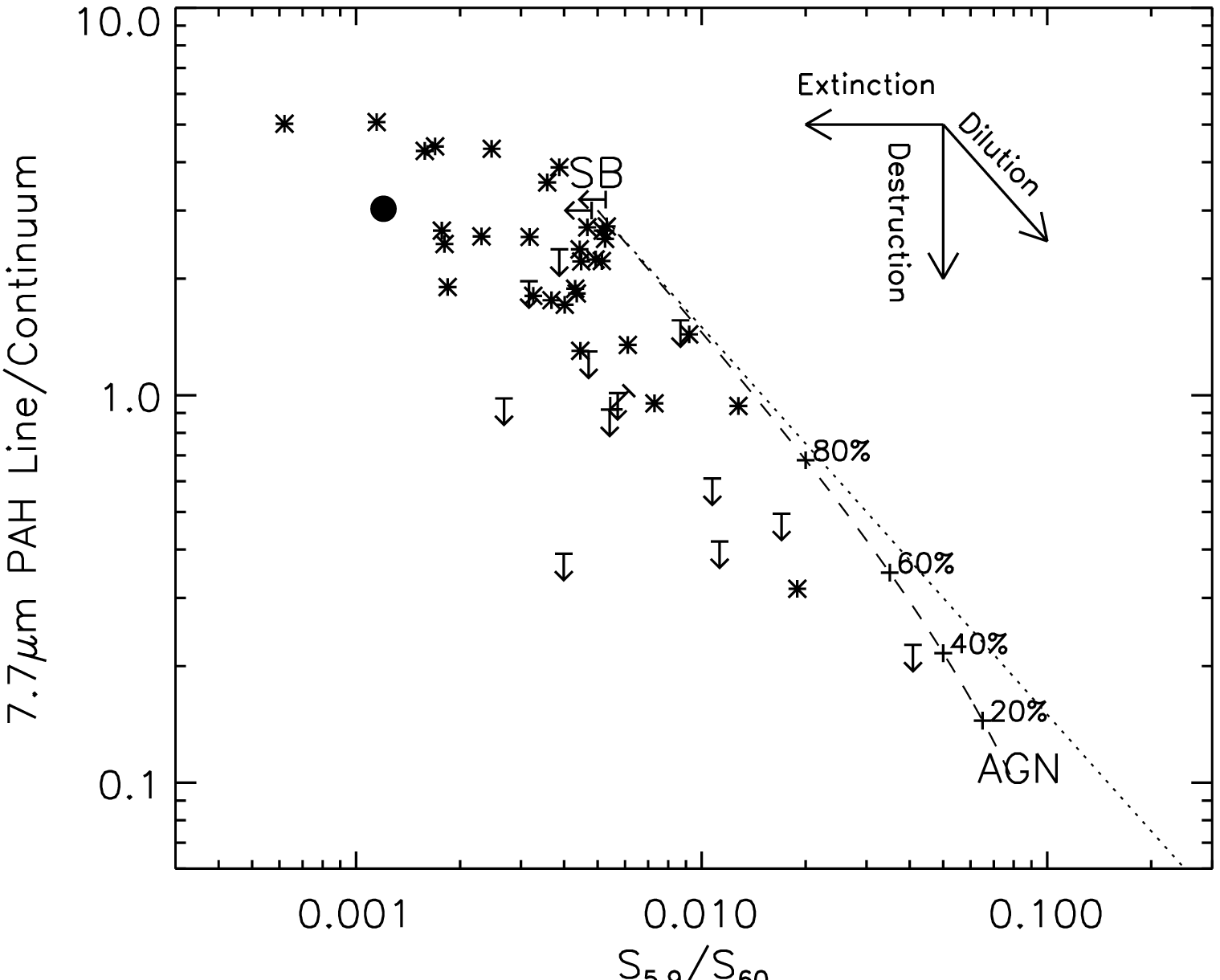]{Line-to-continuum ratio for the 7.7$\mu$m PAH feature 
as a function of the 5.9$\mu$m/60$\mu$m continuum flux ratios. The filled
circle represents the average spectrum of the 14 faint sources with noisy
spectra. Pure dilution of a starburst spectrum by 5.9$\mu$m continuum
(dotted) is indicated, as well as a mixing line (dashed) for the case of  a
starburst ('SB') and AGN contributing different fractions of the 60$\mu$m
flux.
\label{fig:pah_59}}

\figcaption[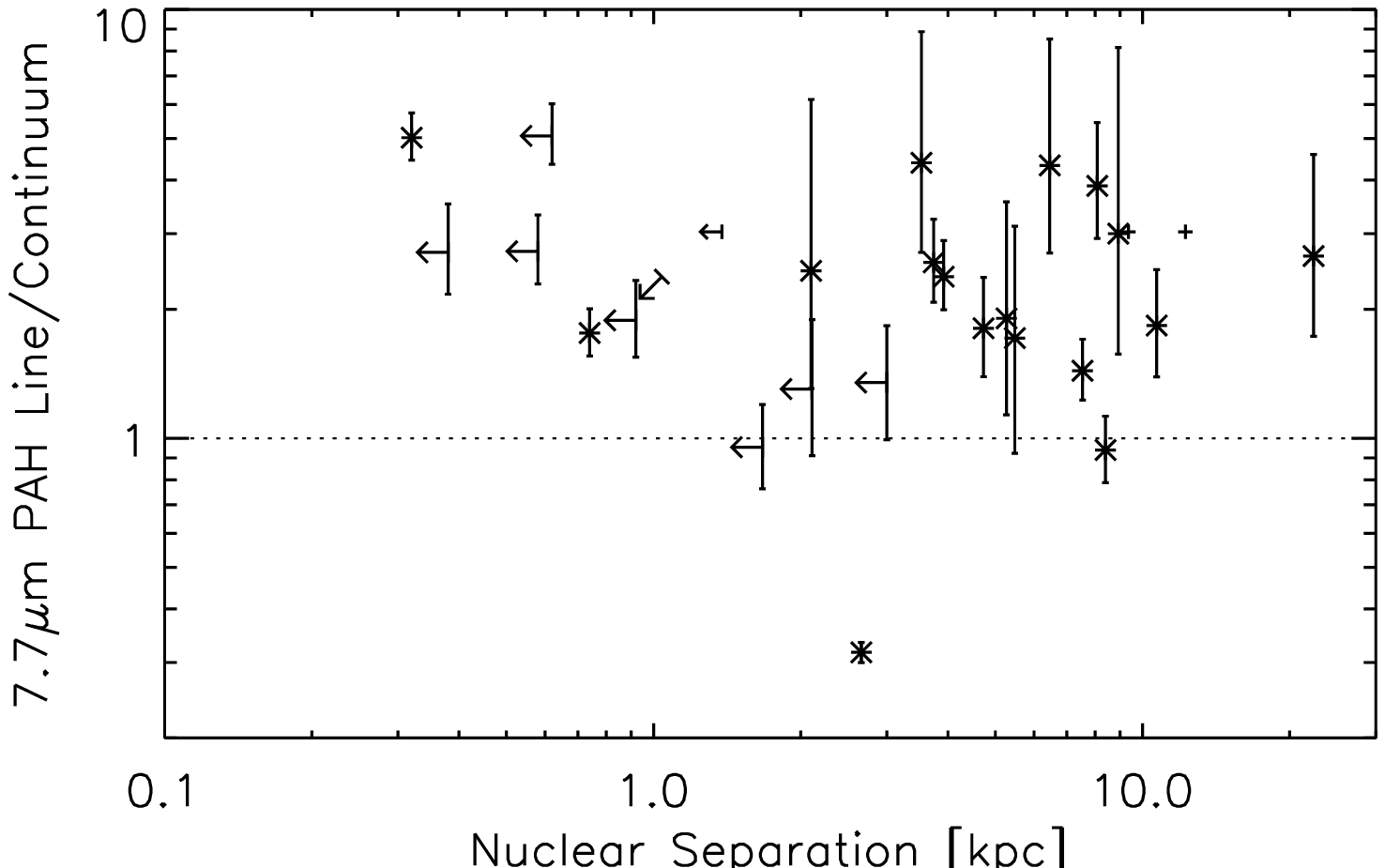]{Line-to-continuum ratio for the 7.7$\mu$m PAH feature 
as a function
of nuclear separation of the interacting components of an ULIRG. We have
placed faint sources with noisy spectra (small symbols) at L/C=3 derived 
from their average spectrum. The dotted line indicates the adopted separation
between starbursts and AGNs at L/C=1.
\label{fig:pah_sep}}

\clearpage
\plotone{pah_meth.eps}
\clearpage
\plotone{pah_lum.eps}
\clearpage
\plotone{pah_59.eps}
\clearpage
\plotone{pah_sep.eps}

\end{document}